\documentclass{article}

\usepackage{wasysym, multicol,graphicx
}
\begin{document}
\def\thebibliography#1{\section*{REFERENCES\markboth
 {REFERENCES}{REFERENCES}}\list
 {[\arabic{enumi}]}{\settowidth\labelwidth{[#1]}\leftmargin\labelwidth
 \advance\leftmargin\labelsep
 \usecounter{enumi}}
 \def\newblock{\hskip .11em plus .33em minus -.07em}
 \sloppy
 \sfcode`\.=1000\relax}
\let\endthebibliography=\endlist

\hoffset = -1truecm
\voffset = -2truecm


\title{\large\bf
Duality : A Bridge Between Physics And Philosophy?\footnote{  This article is dedicated to the memory of Prof D.S. Kothari on his 
birth centenary. }}
\author 
{\normalsize\bf
A. N. Mitra \thanks{e.mail : ganmitra@nde.vsnl.net.in}
}

\maketitle

\begin{abstract}
A generalized view of Duality is offered as a bridge between physical sciences and the more abstract philosophical dimensions 
bordering on mysticism. To that end several examples of duality are first cited from from conventional physics sectors   to illustrate the 
obvious powers of this principle. These include items from reciprocity in Newtonian mechanics to  the problem of measurement 
duality that characterizes   quantum mechanics. It is also noted that the latter  has acquired a renewed  interest in recent times,   
consequent on the emergence of new experimental techniques  for testing the actual laboratory outcomes of  traditional gedanken  
experiments, hitherto taken for granted. Against this background, the Duality principle is  sought to be extended to the mystical domain, with 
convincing  examples from various human  level experiences.  
\end{abstract}
                                                    
\section{Introduction} 

The year 2005  that  has just ended, witnessed the centenary  of Einstein's  1905 revolution  that changed the entire face of Physics. 
And now comes 2006 which is of special significance to this country, namely,  the birth centenary of D.S. Kothari, a  philosopher  
scientist  who  played a key role in directing  the course of science in Independent India, by stressing  the "value system"  that 
goes with it. Therefore in tune with his value-based philosophy, it is  perhaps appropriate to choose the theme of this article as Duality, since  
it  touches  on both aspects of knowledge, namely  physics and philosophy.  The idea is not completely new, since I had written a similar 
article 20 years ago, on the occasion of his eightieth birthday [1].  While the theme of the present article overlaps with that of [1], and  
I shall draw freely from [1],  the relative emphasis   nevertheless  warrants a  fresh  presentation  of the Duality theme.  
 \par
Now  the traditional implication of Duality in physics concerns the incompatibility of measurement of two canonically conjigate variables. 
However, in  tune   with [1], it is  both necessary and sufficient to give an extended meaning of this magic word to cover several other  
sectors  of physics as well.  To that end, we identify as many as FIVE  distinct  facets of duality in physics, namely: Reciprocity;  
\quad Parallelism; \quad Alternative formulation; \quad Synthesis;  and of course Measurement incompatibility.  

\section{Aspects of Duality in Physics}

We  give in this Section some illustrative examples of the first four kinds of Duality in Physics, leaving the (more involved) 
fifth one for the next Section. 

\subsection{Reciprocity (Mutuality) Aspect}

For certain situations, the mathematical equations suggest a sort of mutuality or reciprocal relationship between the dual partners. We 
list three examples [1], each of which implies a deep underlying symmetry in the corresponding physical situation.  The first relation is   
$$  {\vec  A} (action)   =  - {\vec R} (reaction)  $$ 
The symmetry implied by this relation is that the mutual potential energy of any two bodies is an invariant  function of their relative distance,  
and $not$  of their absolute positions.   The next one is from  Maxwell equations in vacuum : 
$$ {\bf \nabla}\times {\bf E} = - \partial_t {\bf B}/c  ; \quad {\bf \nabla}\times {\bf B} = \partial_t {\bf E}/c  $$ 
 Such a reciprocal relationship between the electric and magnetic vectors was conjectured by Maxwell to bring out the full symmetry of their mutual interdependence, which in turn is essential for  current conservation, while the tiny `displacement current' (detected later)  was a byproduct of 
Maxwell's  deep insight into inter-relationship between the two vectors.  A third relationship of this kind stems from the Hamilton's equations 
of motion in terms of canonically conjugate variables $(p, q)$ :
$$ \partial_p H = d q /d t ; \quad \partial_q  H = d p / d t $$   
These equations, whose physical content is strictly confined to Newton's $original$ laws of motion, bring out rather convincingly the reprocity 
of the roles of the $q$- and $p$-variables, a feat achieved through Hamilton's penetrating formulation of the laws of Newtonian mechanics, which 
was subsequently to pave the `golden road to quantization' [2] at the hands of Heisenberg and Schroedinger. 
\par
More examples of mutually inter-dependent pairs may be found from the mathematical theory of transforms (Fourier, Hilbert) which reveal such relationships (together with their diverse physical consequences) in a most succinct manner. Thus while the theory of Fourier transforms is at the root of duality between coordinate space $)x)$ and wave number space $(k)$, Hilbert transforms illustrate the value of analyticity  in the complex plane in 
bringing out the inter-relationship among the real and imaginary parts of a scattering amplitude. There are many such duality relationships in physics.  

\subsection{Parallelism (Analogy) Aspect}

The analogy aspect is best illustrated by Fermat's principle for optics, versus Maupertius principle for mechanics :
$$ \delta \int \mu ds =0  \Leftarrow  \Rightarrow  \delta \int p ds =0 $$ 
This close parallelism between the respective laws between two widely different branches of physics was to play a crucial role in Schroedinger's formulation of wave mechanics from its classical "ray" ($\hbar \rightarrow 0$) picture, as the mechanical analog of wave vs ray optics.   
\par
A brilliant example of the parallelism aspect of Duality is expressed by a profound correspondence between classical and quantum mechanics  
in the form 
$$ \{A, B\} \Leftarrow \Rightarrow \frac{1}{i \hbar} [A, B]  $$ 
discovered bt Dirac during one of his long evening walks in his early Cambridge days [3] 

\subsection{Alternative Formulation Aspect}

Still another feature of Duality concerns the formal equivalence of certain alternative formulations apparently unrelated to each other, yet having the same physical content. The Heisenberg vs Schroedinger formulations of quantum mechanics represent precisely such a physical situation.  Although their equivalence is now text-book material, their apparent dissimilarity at the initial stages of formulation had catalyzed the polarization of two strong schools of thought [4] , viz., i) Heisenberg's algebraic approach emphasizing the corpuscular aspect characterized by discontinuity; vs ii) Schroedinger's analytic approach in terms of a wave equation stressing the elements of continuity [1].  Subsequently it was found that the two approaches differed only by a 
unitary transformation ! 
\par 
A  dramatic manifestation  of  the alternative formulation aspect  showed up in the orthodox  Tomonaga-Schwinger (field theoretic)  versus  
the highly unorthodox Feynman (diagrammatic)   formulations of  covariant quantum electrodynamics. The two formulations looked widely different, but 
it  was Dyson's catalytic role  which reconciled  the two  when he [5] derived the  Feynman diagrams from the premises of  the Tomonaga-Schwinger formalism. And later,  the traditional rivalry between Feynman and Schwinger showed up even more acutely in terms of Path Integral formalism [6]  of the former,  versus Source Theory [7] of the latter, even though the respective contents are  basically  identical ! 
\par
At a more impersonal level, a good example of the complementary aspect of Duality is afforded by the empirical finite energy sum rules (FESR), wherein the contributions to a high energy scattering amplitude by the direct s-channel resonances are supposed to saturate the corresponding contributions from the exchange ($u, t$) channels [8].  This form of duality led to the Veneziano model [9] which received considerable refinements at various hands 
(Harari, Rosner, Fubini et al), eventually giving rise to Nambu's String Model [10], that  was the forerunner of the modern theory of strings.  
 
\subsection{ Synthesis (Unification) Aspect} 

A major aspect of Duality, relating to the synthesis of certain pairs of physical concepts, is embodied in the theory of relativity which provides an integrated view of the space-time continuum, as opposed to the Newtonian `partition' of their respective foundations [1]. The parallel concepts 
of wave-vector \& frequency; momentum \& energy 4-vectors are linked to space-time by Fourier transforms and canonical conjugation respectively. 
On the other hand their direct link is provided by quantum theory : 
$$  {\bf p} = \hbar {\bf k};  \quad E = \hbar \omega $$     
The situation is illustrated in Fig 1. 
\begin{figure}
\caption{A three-way interlinkage}
\includegraphics{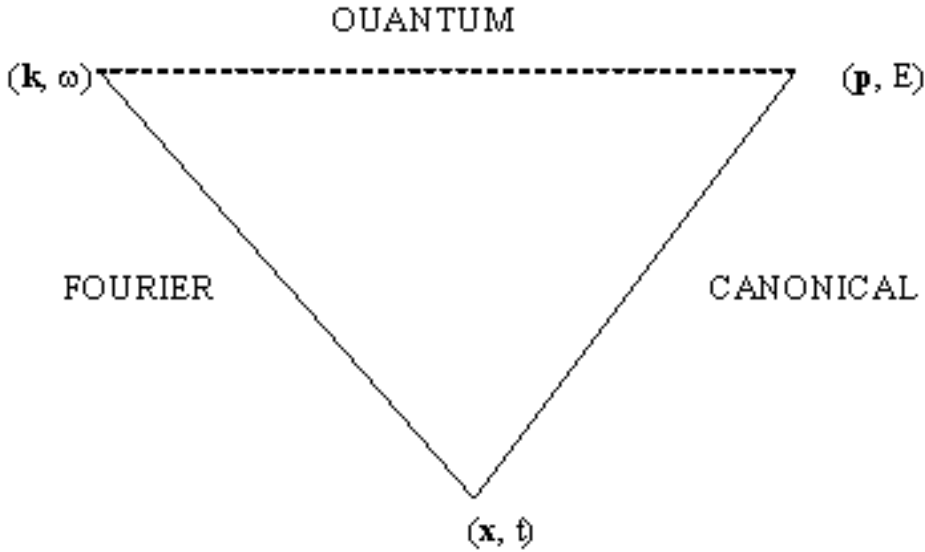}
\end{figure}
\par
A dramatic manifestation of the unification aspect of Duality is the prediction of $antimatter$ as a result of the successful marriage of the dual 
partners represented by Relativity and Quantum theory under the auspices of Dirac who effectively showed that the marriage is not possible 
at the level of single particles, but only in the collective context of particles and antiparticles; in other words, in a $field$ theory [11]. 
\par
As a final example of the unification aspect of Duality, the concept of $Supersymmetry$ [12]  purports to project an integrated view of $bosons$ 
and $fermions$, hitherto believed to be two distinct fundamental species, totally unrelated to each other. The Bose-Fermi symmetry or SUSY, as this new theory is called, has some highly attractive theoretical features, like an ability to cure some vexing problems of infinities, but its predictions are yet to find experimental support. The theoretic investment in this field has been extremely rich in recent years, with allied developments in Supergravity,  superstrings, and so on, but there has been almost zero development on the (dual) experimental front.    
Table 1 below gives a list of the various dual partners in physics, discussed in the foregoing. 

\begin{table}
\caption{Dual partners (physics items)}
\begin{tabular}{p{4cm}p{4cm}p{4cm}}
\hline \\
{\it I-Member} 				& {\it II-Member} 			& {\it Legend/Ref.} \\ 
Action ($A$) 				& Reaction($R$)			& Newton III \\
Coordinates ($q, \theta$)         	& Momenta ($p, J$)       	            & Hamilton, Jacobi \\
Electricity				& Magnetism				& Faraday--Maxwell \\
{\bf E}, {\bf B}				& {\bf D},{\bf H}			& Cause vs effect \\
Pressure, stress			& Volume, strain			& Cause vs effect \\
Time ($t$)				& Space ($r$)				& Einstein's relativity \\ 
Energy ($E$)				& Momentum ($p$)			& Relativity  \\
Energy ($E$)				& Mass ($m$)				& Relativity \\
Fermat principle			& Maupertius principle		& Parallelism \\
$\delta \int \mu ds = 0$		& $\delta \int \rho ds = 0$		& Optics vs Mechanics \\
electromagnetic wave			& Photon				& Planck \\
$e^{-}$ - wave				& Electron				& de Broglie \\
Schroedinger (wave mech)	 	& Heisenberg (matrix mech)		& Alternative formulations \\
Feynman   path integral	            & Schwinger  Source theory                & Language duality \\
Resonances  (s-channel)	            & Exchanges (t u channel)                  & FESR duality ( ref. [8] ) \\                                                                                                                        
Observer ( macro apparatus)		& Observable (micro system)	            & Bohr's duality \\
Boson					& Fermion				& Supersymmetry \\ \hline
\end{tabular}
\end{table}

\section{Measurement Aspect of Duality}

Finally we come to the most important (quantum mechanical) aspect of Duality, namely, the incompatibility of measurements of certain pairs of dynamical variables known as canonically conjugate pairs, together with their derivatives.  This shows up as the famous Uncertainty Principle of Heisenberg, which is mathematically derivable from any consistent form of quantum mechanics (Heisenberg, Schroedinger), in the form:
$$ \Delta q .\Delta p \geq \hbar / 2 ; \quad if \quad [q,p] = i\hbar $$   
Although a strict mathematical consequence of the tenets of quantum mechanics, the physical significance of the Uncertainty Principle (UP)  is nevertheless profound enough to touch instantly on the philosophical plane. For one thing, it succinctly reveals the paradox of the wave-particle duality which is profoundly disturbing inasmuch as it goes against all norms of "classical justice". For another, the nature of a `gedanken experiment' by which the effects of the  UP is sought to be  projected,  happens to be  such as to preclude the possibility of simultaneous observation of two canonically conjugate variables in the $same$ experiment . Indeed,  any attempt to measure the second variable in an experiment designed to observe the first, will result in a destruction of the property of the former, and vice versa !  This limitation transcends either the quality of the experimental apparatus or the extent of the human ingenuity in designing the experiment, since it stems directly from the mutual interaction between the observer (apparatus) and the observable ( the physical property under study). To make contact with the corresponding `classical' situation, the effect of this interaction is negligible on 
a macroscopic entity, but non-trivial on a microscopic one (of atomic dimensions), so much so, that an accurate measurement of one its attributes 
precludes a simultaneous knowledge of the  canonically conjugate one.   
\par
The last aspect of Duality has no counterpart in the other four categories listed earlier, since there had been no reference to the quantum limitation 
in any of them. Now the measurement incompatibility problem is a typical quantum effect, which introduces a characteristic `duality' situation that has  no classical counterpart.  And several of the pairs considered in the foregoing will suffer the  measurement limitation if viewed from  a quantum mechanical   
angle. A good example is afforded by the `reciprocity related' variables ${\bf E, \bf B}$ which indeed suffer the measurement incompatibility problem 
when viewed quantum mechanically, using the idea of `test charges' for their measurement [13]. 

\subsection{The Copenhagen Interpretation}

The Copenhagen Interpretation, which is based on Bohr's view of Duality, is concerned with the problem of observer-observable interaction  at the 
quantum level. In a lecture at the International Physics Congress at Como in 1927 [1], he introduced the principle of $complementarity$ to reconcile 
the characteristic features of individual quantum phenomena  with the observational problem  "in this field of experience". In particular, he emphasized "the impossibility of any sharp separation between the behaviour of atomic objects and their interaction with measuring instruments which serve to 
define the conditions under which the phenomena appear" (quote from Bohr [1]). 
\par
This physical picture, which was Bohr's response to the mathematical content of the UP, was fully in consonance with the deep insight which had 
marked his stewardship of  atomic physics since its nascent beginnings early in the last Century. In contrast, the absolute Wave Function approach 
of Von Neumann [14], whose adherents included  stalwarts like Wigner, Everett, Wheeler and de Witt (see ref [1]), advocated the use of a master wave function which included the wave functions of both the (microscopic) system under observation, and that of the (macroscopic) measuring apparatus.  
Without going into the elaborate rules governing the interaction between these two systems [15], the Copenhagen Interpretation takes an intensely 
pragmatic view of the concept of `truth' ( a la William James ?), as may be summarized in the following two statements: \\
(a) The quantum theoretic formulation must be interpreted pragmatically [1] ; \\
(b) Quantum theory provides  for a  $complete$  $scientific$ account of atomic phenomena. \\
The pragmatic aspect  has found expression in numerous thoughts and writings of Bohr on the measurement process itself (for details, see ref.[15]). 
The `completeness' aspect of quantum theory  is more subtle, and gave rise to the Bohr-Einstein debate.  Bohr's point of view in this regard is aptly 
summarized by Stapp [15].  Namely [1], the well-defined objective specifications on a given phenomenon under study are $not$ restrictive enough to determine uniquely the course of the individual processes, yet no further breakdown is possible because of the inherent $wholeness$ of the process symbolized by $\hbar$. This `wholeness' has no classical analogue which would have recognized the measuring instruments and the atomic objects as separate entities. Instead, the inseparability of the atomic object from the entire phenomenon renders a statistical description unavoidable. This way of reconciling the pragmatic character of quantum theory with the claim of completeness, is based on " quantum thinking". Its ultimate validity must be 
judged by its $afortiori$ success which includes coherence and self-consistency. 

\subsection{ EPR paradox and Bell's Theorem}

What was Einstein's reaction to Bohr's philosophy ? It was one of profound unhappiness [16] with such claims of "completeness" of quantum mechanics !  Compare  the above  picture with the tenets of classical realism, namely,  (a)  all physical attributes of an individual   
object have definite values associated with them at any instant of time, $irrespective$ of their actual ( non-invasive) measurement ; (b)  a commonsense concept $locality$ which allows us to deal with the external world in a piecemeal fashion, not all at once.   Now look at the situation in  quantum mechanics : if two systems have once interacted together, and later separated  (no matter how far), they can no longer be assigned separate state vectors. A famous example is a spin-zero object at rest, breaking up spontaneously into two fragments with spins $S_1$ and $S_2$ respectively, moving in opposite directions. Now conservation of angular momentum demands that the two spins be equal and opposite, so that any measurement of, say, $S_1$ will automatically fix the value of $S_2$, even without an explicit measurement.   This situation goes much against intuition, since a physical interaction between these two objects, receding far away from each other, is negligible. And this was the bone of contention of the  EPR  paper [16] on this paradoxical aspect of quantum mechanics.  EPR [16] sharpened the issue further by introducing two definitions [17, 18] :
 i) a necessary condition for the completeness of a theory is that " every element of the physical reality must have a counterpart in the physical theory " ; ii) a sufficient condition to identify an element of physical reality is " if in any way without disturbing the system, we can predict with certainty the value of a physical quantity, then there exists an element of reality correspondity to this physical quantity ".  The result of these considerations was the EPR Theorem [16], viz., the $incompatibility$ of the following two statements : \\
1) The description via the $\psi$ function of quantum mechanics is complete ; \\
2)  The real states of spatially separated objects are independent of each other.   \\
The second statement goes by the name of " Einstein's locality postulate ", which is clearly incompatible with the first, which asserts that quantum mechanics is a $complete$ description !  This incompatibility is the EPR Theorem.  Thus there is a conflict between classical and quantum realisms, 
to resolve which calls for a precise experimental test. This test was formulated by John Bell [19] who made the concept of Einstein locality more 
precise by introducing "hidden variables" as a means of circumventing the counter-intuitiveness implied in the quantum  description, and formulating suitable experimental tests in the form of Bell's inequalities for a suitable combination, say $F$,  of the amplitudes for two spin-half particles moving in opposite directions.  These combinations would differ according as the information on Einstein locality (no correlation for widely separated particles) , or that for `quantum entanglement' (no matter how much their separation is), is incorporated.  Then Bell's inequality reads as 
$$  | F |  \leq 2 ;  \quad | F |  \geq 2 \sqrt {2} $$ 
for the two cases respectively.  Actual experiment [20] upheld quantum mechanics, thus vindicating Bohr at least for now.   

\subsection{Bohr - Einstein Duality}

The Bohr-Einstein controversy  is a new form of Duality arising  from the measurement problem. This is a strict   consequence  of  the advent of 
Quantum theory which formally marked the demise of the so-called Cartesian Partition between the physique and the psyche, and brought about 
an intricate interaction between the two.  The issue is one which deeply involves physics with philosophy,  but  it was  lying more or less dormant for many decades. And  now it seems to  have suddenly sprung up to life during the last 3 decades, thanks to the i)  growth of new experimental techniques (which have given a fresh lease of life to the experiments hitherto thought to be at the `gedanken' level) on the one hand, and  to the ii) development of quantum technologies of information processing and transfering on the other [21].  The movement  has indeed grown up into a full-fledged industry, and shows  no signs of abating, but its direction is not yet clear. It has no doubt generated a good deal of  heat  in the form of  a glossary of technological jargon, but its concrete success on the physics front is so far  highly questionable, since  a resolution of the actual Bohr-Einstein debate is still far from over.  Nevertheless, the physics-philosophy interaction that has thus been generated, formally opens up a $bridge$ for taking physics to a  higher plane 
of $consciousness$.  The subject is not, since several stalwarts  like  Fritzof Capra [22], have addressed  the issue,  but the renewed interest 
generated by the measurement vis-a-vis locality problem gives it a further push.  And this brings us to the last phase of this paper which offers 
a  glimpse of what lies beyond. 

\section{Duality  Partners Beyond Physics} 

So far we have discussed certain concrete facets of Duality during the historic growth  of physics through the ages since Newton.  In this development,  the Cartesian Partition had remained  in the background, without publicly appearing to influence the `contingent plane' of empirical and analytic statements, a la Holton [4]. Newton himself had been aware of the duality between the physique and the psyche, but was inclined to project only the former, without active encouragement to the latter. His predecessor Kepler, on the other hand, had relied more heavily on the thematic concepts of the universe as a `mathematical harmony',  and a central theological order [4]. And Newton's decisive influence on Western scientific thought had much to do with the uneasy balance between a materialistic pursuit of science and an idealistic devotion to philosophy, that had characterized the thematic development of physics till the early part of the twentieth century, when Einstein and Bohr came on the scene [23].   Einstein's deep philosophy behind his unified view of space-time continuum on the one hand, and Bohr's physical insight leading him to enunciate the Complementarity Principle (CP) as a paraphrase of the Uncertainty Principle on the other, marked such a radical departure from the Western attitude to science prevalent till then, that these had the effect of a "wind of change" on a relatively close and still atmosphere.   In particular, the CP  set the Western community of physicists on the formidable task of reorienting their attitudes as a result of intrusion of dialectics into their traditional modes of thinking. Interestingly enough, Bohr's exposure of the same philosophy before the Japanese community met with little resistance to their traditional Eastern thought , as recounted by Yukawa to Rosenfeld [23].    
\par
What is the extra ingredient in Eastern thought with which Bohr's CP philosophy found such a ready resonance , despite sounding so unorthodox to the Western school ?  This brings us to the contents of  Table 2 which lists some dual partners on the interface between  physics and philosophy that cannot be fathomed with the `standard'  methods of physical science. The items listed in this table need to be read from intuitive and commonsense considerations,  with apologies to the  orthodox methods of physical science (see below for further comments).  
\par
For any science in its formative stages, the traditional methods of limited hypothesis, checked against vigorous experimentation and vice versa, 
have usually proved much more effective than unfettered speculation of ideas with no comparable degree of experimentation to provide the balance. 
There comes nevertheless a stage in its development when this relatively mundane method fails to do adequate justice to the intellectual aspirations of the scientific thinker. A very similar stage has been reached in modern physics where the unification of opposite concepts represents precisely such an aspiration where the experimental support often lags so far behind the theoretical ideas, that $faith$ in the latter must, in the interim, be sustained by considerations of a thematic nature, well before eventual experimental confirmation, if at all. Such opposite concepts abound at the sub-atomic level where particles are both continuous and discontinuous; and force and matter are but different aspects of the same phenomena. In all these examples it turns out that the " framework of opposite concepts, derived from our everyday experience, is too narrow for the world of subatomic particles " (Capra, ref [22]).    
\par 
Some of these situations have already been illustrated under the unification (as well as measurement incompatibility ?)  aspects  of Duality in the foregoing. In each case, the unification occurs on a higher plane; e.g., space and time become a single entity only in a 4-dimensional continuum; 
wave and particle manifestations of an electron / photon get unified only at the quantum level;  matter and anti-matter require a further synthesis of relativity and quantum theory for a mathematically self-consistent description ; and so  on.   Fritjof Capra in his remarkable book, Tao of Physics [22], 
has documented a large class of such examples (through extensive quotations from appropriate religious, philosophical and scientific authorities), 
in his exploration of parallels between Western physics and Eastern mysticism, and revealed a profound harmony between the two.  In particular, he portrays a simple picture, emanating from the Chinese symbolism of the archetypal poles $Yin$ and $Yang$-- two extremes that are constantly engaged in a dynamic interplay which brings about their unity ($Tao$) on a higher plane.  This picture has a simple physical  analog in the example of a circular motion and its linear projection [22], which is illustrated in Fig. 2. Note that the continuous oscillation between the two opposite points ($Yin$ - $Yang$)
is a characteristic only of the linear projection, while the more complete two - dimensional circular motion shows no such fluctuation.             

\begin{table}
\caption {Dual partners (beyond physics)}
\begin{tabular}{ll}\hline
Theory (interpretation)		            & Experiment (data) \\
Subject				& Object \\
Subjective conjectures		& Objective analysis \\
Mind (consciosness)			& Brain (nerve complex) \\
Psyche 				& Physique \\
Religion (philosophy)			& Science (pragmatism) \\
Heart (emotion)			& Head (reason) \\
Mysticism (eastern)			& Modern science (western) \\
Purusha (Source of energy)		& Prakriti (Manifestation of energy) \\
Bhakti (blind faith)			& Gyan (reasoned knowledge) \\
Yin (creative instinct)		            & Yang (mental exercise) \\
Faraday--Bohr (intuitive flash)	& Maxwell-Dirac (mathematical precision) \\
Private science (random ideas )	& Public science (expository skills) ref [4] \\
Simplicity 				& Thoroughness  ( Bethe, ref [1])  \\  
Dreaming				& Doing \\ \hline
\end{tabular}

\end{table}

\begin{figure}
\caption{The dynamical synthesis of opposites}
\includegraphics{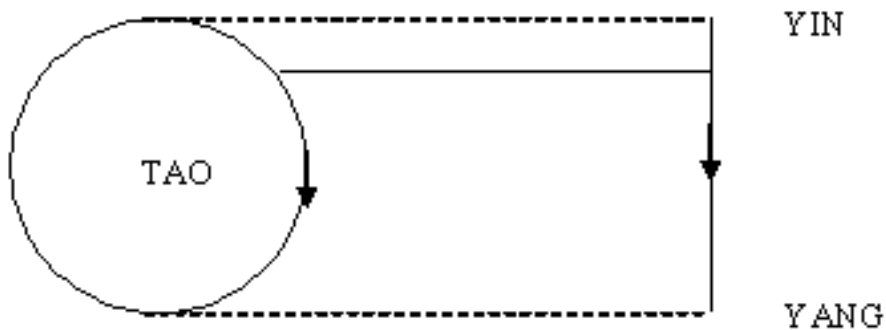}
\end{figure}
      
\par
At this stage,   it should be appropriate to cite an example given by the Man  in whose honour  this article has been  designed.  And this is  from the Jain Philosophy of Saidavada which shows a  remarkable parallel in  wave-particle duality of quantum mechanics, via its $fourth$ mode of realization of reality, namely $ Avayakta$  (inexpressibility), as  described by Kothari  in one of his last publications  [24].      
  
\subsection{  Consciousness  Dominated Physics ?}

Since consciousness is very much a part of the physique - psyche duality,  one might wonder  if  the roles of these two items could be  interchanged, and  a  quantum theory  based on the dominance of  the  psyche (consciousness), instead of the more conventional theory dominated by the physique (matter), be  pursued seriously.  Actually  one such approach has been offered in   a recent  book [25] by   Amit Goswami, a  physicist- turned- philosopher, 
but it  is still too premature to let such ideas, howsoever appealing,  compete on equal terms with conventional quantum mechanics.  

\section{Conclusion}

Coming back to our Duality theme, we end this narrative with the observation that mystics through their meditation transcend the realm of intellectual concepts and in so doing, become aware of the polar relationship of all opposites. Physicists grope for a glimpse of the same through their language of mathematics. Western philosophers have been keenly aware of this duality ( see, e.g.,  Emerson's thesis on the hidden law of compensation [1]). And today, theoretical physicists are increasingly feeling its impact as they probe deeper into the mysteries of the sub-atomic world down to $10^{-17} cm$, but theoretically all the way to Planck's length, having received its first taste in Bohr's CP, supported experimentally by wave-particle duality. There is no going back on this journey , irrespective of the source, be it modern physics or mysticism, of its inspiration.



\begin{thebibliography}{99}
\bibitem{1}
A.N. Mitra, {\it Duality in Physical Sciences and Beyond}, Pramana-J. Phys.{\bf 27}, 73-87, (1986).  
\bibitem{2}
H. Goldstein, {\it Classical Mechanics}, Addison-Wesley, Reading, MA, USA, (1950). 
\bibitem{3}
P.A. M. Dirac, in {\it A Lifetime of Physics}, IAEA , Vienna, (1968) 
\bibitem{4}
G. Holton,  {\it Thematic Origins of Scientific Thought}, Harvard U Press, (1973). 
\bibitem{5}
F.J. Dyson, Phys. Rev.{\bf 75}, 486, 1736 (1949). 
\bibitem{6}
R.P. Feynman and A.R. Hibbs, {\it Quantum Mechanics and Path Integrals}, McGraw Hill, N.Y., (1965). 
\bibitem{7}
J. Schwinger, {\it Particles, Sources and Fields}, Addison Wesley, Reading MA, (1973).  
\bibitem{8}
R.Dolen et al, Phys. Rev.{\bf 166}, 1768 (1968); \\
A.A. Logunov et al, Phys. Lett. {\bf B24}, 181 (1967). 
\bibitem{9}
G. Veneziano, Nuovo Cimento {\bf 57}, 190 (1968) 
\bibitem{10}
Y. Nambu, in {\it Symmetries and Quark Models}, (ed) R. Chand, Gordon and Breach, N.Y. (1970). 
\bibitem{11}
F.J. Dyson, {\it Lecture Notes on Advanced Quantum Mechanics}, Cornell Univ,  (1952). 
\bibitem{12}
J. Wess and B. Zumino, Nucl.Phys.{\bf B70}, 109, (1974) 
\bibitem{13}
N. Bohr and L. Rosenfeld, in {\it Developments in the Theory of the Electron}, (ed) A. Pais, Princeton U Press, (1948). 
\bibitem{14}
J. von Neumann, {\it Mathematical Foundations of Quantum Mechanics}, Princeton U Press (1955). 
\bibitem{15}
See, e.g., H.P. Stapp, Am J. Phys.{\bf 40}, 1098 (1972). 
\bibitem{16}
Einstein, Podolsky and Rosen, Phys. Rev.{\bf 47}, 777 (1935) 
\bibitem{17}
V. Singh, quant-ph/0412148. 
\bibitem{18}
A.N. Mitra, quant-ph/0510223 
\bibitem{19}
J. S. Bell, Physics {\bf 1}, 195 (1964). 
\bibitem{20}
A. Aspect et al, Phys. Rev. Lett. {\bf 49}, 91, 1804 (1982). 
\bibitem{21}
D. Home, {\it Perspectives in Quantum vs Classical Reality}, 2002 (unpublished). 
\bibitem{22}
F. Capra, {\it The Tao of Physics},  Fontana,  London (1976).
\bibitem{23}
L. Rosenfeld, Phys. Today{\bf 16}, 47, (1963). 
\bibitem{24}
D.S. Kothari, {\it Complementarity and Syadavada} in {\it Niels Bohr--A Profile}, (ed) A.N. Mitra et al, Ind Natl. Sci. Acad, New Delhi, 1985.  
\bibitem{25}
Amit Goswami, {\it Physics of the Soul}, Hampton Roads Pub Co, Charlotsville VA, 2001. 


\end{thebibliography}
\end{document}